\documentclass[onecolumn,letterpaper,11pt,draftclsnofoot]{IEEEtran}
\usepackage{amssymb}
\usepackage{amsmath}
\usepackage{amsthm}
\usepackage{amsfonts}
\usepackage{mathrsfs}
\usepackage{float}
\usepackage{graphicx}
\usepackage{epsfig,graphics,subfigure,psfrag}
\usepackage{pdfpages}
\usepackage{enumerate}
\usepackage{engord}
\usepackage{cite}
\usepackage{verbatim}
\usepackage[ruled]{algorithm2e}
\usepackage{url}

\begin{document}

\newtheorem{definition}{\bf Definition}
\newtheorem{theorem}{\bf Theorem}
\newtheorem{lemma}{\bf Lamma}
\newtheorem{proposition}{\bf Proposition}

\title{Game Theoretic Approaches to Massive Data Processing in Wireless Networks}
\author{\normalsize{\IEEEauthorblockN{{Zijie Zheng}}, \IEEEmembership{\normalsize{Student Member, IEEE}},
\normalsize{Lingyang Song}, \IEEEmembership{\normalsize{Senior Member, IEEE}},\\
\normalsize{Zhu Han}, \IEEEmembership{\normalsize{Fellow, IEEE}},
\normalsize{Geoffrey Ye Li}, \IEEEmembership{\normalsize{Fellow, IEEE}}, \normalsize{and H. Vincent Poor}, \IEEEmembership{\normalsize{Fellow, IEEE}}}\\
\vspace{-0.5cm}
\thanks{Z. Zheng and L. Song are with School of Electrical Engineering and Computer Science, Peking University, Beijing, China (email: \{zijie.zheng, lingyang.song\}@pku.edu.cn).}
\thanks{Z. Han is with Electrical and Computer Engineering Department and Computer Science Department, University of Houston, Houston, TX, USA, and also with the Department of Computer Science and Engineering, Kyung Hee University, Seoul, South Korea. (email: hanzhu22@gmail.com).}
\thanks{Geoffrey Y. Li is with School of Electrical and Computer Engineering, Georgia Institute of Technology, Atlanta, GA, USA (email: liye@ece.gatech.edu).}
\thanks{H. V. Poor is with the Department of Electrical Engineering, Princeton University, Princeton, NJ, USA (email: poor@princeton.edu).}
}
\maketitle

\thispagestyle{empty}
\begin{abstract}
Wireless communication networks are becoming highly virtualized with two-layer hierarchies, in which controllers at the upper layer with tasks to achieve can ask a large number of agents at the lower layer to help realize computation, storage, and transmission functions. Through offloading data processing to the agents, the controllers can accomplish otherwise prohibitive big data processing. Incentive mechanisms are needed for the agents to perform the controllers' tasks in order to satisfy the corresponding objectives of controllers and agents. In this article, a hierarchical game framework with fast convergence and scalability is proposed to meet the demand for real-time processing for such situations. Possible future research directions in this emerging area are also discussed.
\end{abstract}

\section{Introduction}
Due to the development of technologies in mobile sensing, communications,
and storage, mobile data is being generated at unprecedented
rates in wireless communication networks~(WCNs). The global mobile data traffic reached $7.2$ exabytes~($7.2\times10^{18}$ bytes) per month at the end of 2016, up from 4.4 exabytes/month at the end of 2015. With a $47\%$ annual growth rate, the mobile data traffic
is predicted to increase sevenfold and reach 49.0 exabytes/month by 2021~\cite{Cisco}. This information explosion motivates the need for substantial increases in the computing, storage, and transmission capabilities of the devices for data processing. The central network needs to evolve high performance servers to handle diversified multimedia sources and deal with massive data. The base stations~(BSs) or access points~(APs) at the edge need fast processers to handle intensive forward and backward signal processing and operate efficient spectrum, power, and antenna resource allocations. Mobile devices need to be able to upgrade CPUs, memories, and communication units quickly to meet new requirements.\par
\subsection{Hierarchical Network for Big Data Processing}
However, even with a very rapid development of data processing capabilities of servers, processers, and mobile devices, the data processing still lags behind the explosive rate of data generation when each device deals with its data on its own~\cite{Bigdatamag2}. Thus, WCNs are becoming highly virtualized with typical two-layer hierarchies, in which controllers at the upper layer with duties to achieve can ask the nodes at the lower layer as agents to help realize the computation, storage, and transmission functions~\cite{Bigdatamag}. Through dividing the data processing into multiple groups and offloading them onto multiple agents, the controllers can accomplish the otherwise prohibitive big data tasks. Fig.~1 illustrates the hierarchical nature of WCNs at their various levels, i.e., the architecture level, the hardware level, the software level, and the algorithm level, some salient features of which are described in the following.
\begin{figure}[!tbp]
\centering
\includegraphics[width=6.3in]{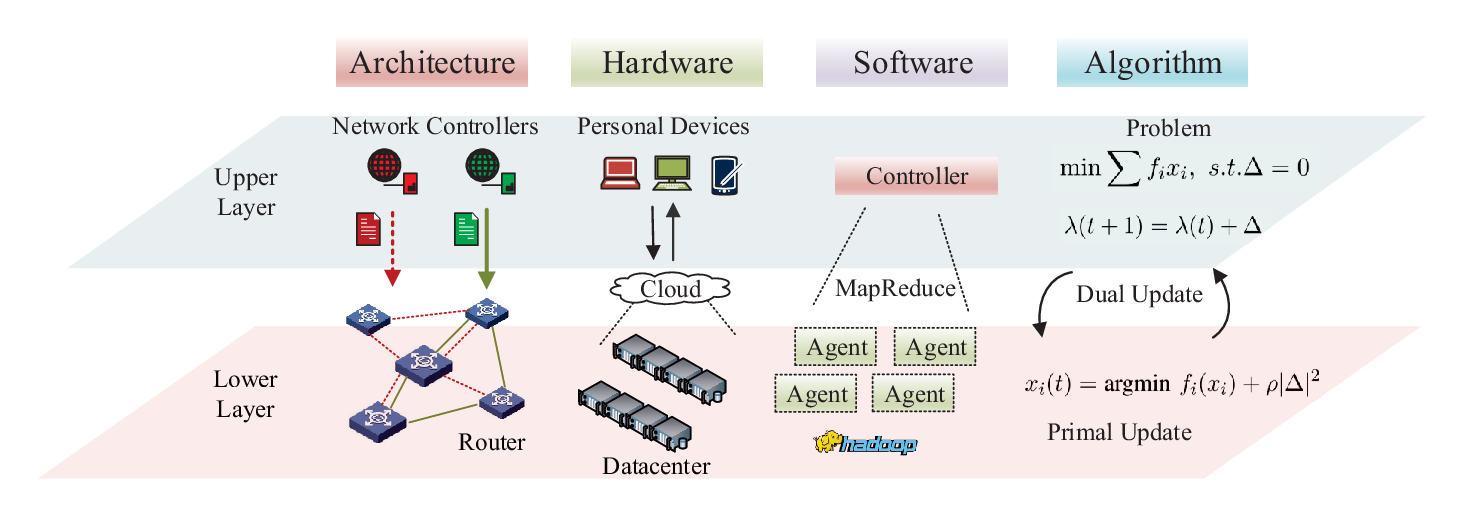}
\vspace{-5mm}
\caption{Hierarchies at the architecture level, the hardware level, the software level, and the algorithm level.}
\label{fig:fig1}
\vspace{-5mm}
\end{figure}
\begin{itemize}
\item \textbf{Architecture Reorganization:}  WCNs are being equipped with more programmable units, replacing the conventional core networks with hardware-based dedicated devices. Different controllers design the software programs for various functions and reuse the same hardware to back up different practical processes~\cite{Architecture}. For instance, the routers/mobile relays can be reused by various controllers to construct
multiple virtualized networks for multiple routing assignments. Similar idea also helps reorganize
the edge network, such as in cloud-based radio access networks~(C-RANs), in which the BSs can
share the same baseband processor with powerful computing capacity to overcome the computation
limitations.  Furthermore, mobile devices held by users and ubiquitous sensors now can share their computing resources with each other with the help of fog computing and Internet-of-Things~(IoT).
\item \textbf{Hardware Deployment:} When the data processing requirements exceed the capacity of a device, it is natural to find more devices to help with computations. By the end of 2016, more than 300 hyperscale datacenters~(DCs) have been built by industrial corporations, such as Google and Amazon. Each DC can support cluster computing with thousands of servers and large capacity memories. Service providers~(SPs), network operators, and even network users can rent several parts of the servers in the hyperscale DCs to accomplish the tasks that cannot be undertaken with their own capabilities.
\item \textbf{Software Development:}
     Since the Hadoop distributed file system~(HDFS) and MapReduce framework were proposed in 2005, software has been developed rapidly to organize the increasing number of hardware devices~\cite{Hadoop}. The HDFS has a hierarchical controller-agent architecture, where the controller can allocate the tasks to agents. When the tasks are finished, the agents return the results to the controller. Similar systems, such as Spark~(2009) and Storm~(2014), have been widely adopted by industries to manage their files and everyday computing tasks.
\item \textbf{Algorithm Design:}
     Besides the well-designed network architecture, hardware, and software, specific distributed algorithms are in a great need to guarantee the convergence for data processing with multiple devices. After modeling data processing as an optimization problem, convex optimization algorithms, such as coordinate descent algorithm, proximal gradient method, network Newton method, and alternating direction method of multipliers~(ADMM)\cite{ADMM}, are very useful to decouple the original problem into different parts to be processed on multiple devices. More importantly, these algorithms can be analytically proved to converge, which reduces time consumed on checking the convergence in practical implementations.
\end{itemize}
\par
\subsection{Game Theory based Methods}
With ubiquitous hierarchies in WCNs, any entity, an SP, a network operator, or a user, can act as a controller and accomplish the data processing tasks with the agents' assistance. However, it is too idealistic to assume that the agents will always help the controller freely. In reality, the agents usually have their own tasks for their data processing resources. The assistance for achieving the controller's objective may jeopardize the original benefits/individual objectives of the agents. Thus, the controller needs to incentivize the agents to compensate loss. Naturally, game theory can be adopted as a promising solution in quantitatively modeling and solving the incentive mechanism design problems~\cite{Game}. For example, in the Stackelberg game for a WCN with one controller and multiple agents, the controller is usually defined as the leader with the objective described by a utility function and the individual objective for each agent is quantified as a follower's objective function. The controller can adjust the payments as its strategies whereas the agent can provide the corresponding data processing resources based on the payments. The game is played iteratively with the payment and resource adjustments until reaching a certain balance/agreement, such as a Stackelberg equilibrium that can satisfy both the controller and the agents. \par
Although the existing game theoretic methods have found
some ways to design incentive mechanisms, the following questions are still open, which restrict the application of the game theory in wireless big data processing~\cite{Game}.
\begin{itemize}
\item \textbf{Convergence Conditions:} When the amount of data from the controllers is large, the number of assistant agents is usually large. In this case, it is difficult to guarantee the convergence of the payment negotiations and resource allocation. A systematic framework is required to clearly point out the relation between the convergence of the games and the forms of utility functions for both the controllers and the agents. Some frameworks have analytically investigated the convergence conditions, including potential games, auctions, mean-field dynamic games, some of the contract games, and coalition formation games. However, they require specifically constructed functions, such as potential functions in the potential games and fixed valuations for commodities in Vickrey-Clark-Groves~(VCG) auctions~\cite{Game}.
\item \textbf{Convergence Speed:} Rapid convergence of the game is a prerequisite in wireless big data processing. Using excessive time for payment negotiations between the controllers and the agents is intolerable for the following two reasons. First, hundreds of iterations between each pair of controller and agent can cost billions of signaling packets overhead when the big data processing tasks need to be allocated to a large number of agents. This may exceed the network transmission capacity. In addition, in some kinds of networks, such as in vehicular networks and industrial IoT networks, the mobility of the devices can only support an ephemeral cooperation among them, which cannot support time-consuming mechanisms.
\item \textbf{Scalability:} Scalability is hard to guarantee in variety of game theoretic approaches, such as the Merge-Split algorithm in coalition games~\cite{Game}. Intuitively, without properly designed mechanisms, the number of iterations will prohibitively increase to balance intertwined objectives from various controllers and agents.
\end{itemize}
\par
\subsection{Contributions}
In this article, we propose a generalized, rapidly convergent, and scalable hierarchical game to provide fast payment negotiations and resource adjustments for the ubiquitous hierarchies for wireless big data processing and in consideration of the above limitations in the existing game theory. Specifically, we design a hierarchical game in three different forms to cope with the networks with one controller and multiple independent agents, with conflicted agents, and with multiple controllers, respectively. The fast convergence of our hierarchical game in each form is guaranteed for a general class of objective/cost functions for the controllers and the agents, i.e., convex functions. Furthermore, the hierarchical game is scalable since the convergence speed is independent of the network size. In brief, our framework can be implemented as a generalized incentive mechanism to help the controllers allocate the big data processing tasks to a large number of agents. \par

In the rest of this article, we first present four typical scenarios in wireless big data processing with explicit hierarchies in Section~\ref{sec:example}.
Then, our hierarchical game is provided in Section~\ref{sec:hierarchicalgame}.
Finally, we present some interesting research topics in this area and conclude the article in Section~\ref{sec:future} and Section~\ref{sec:conclusion}, respectively. \par

\section{Incentive Problems in Wireless Hierarchies}
\label{sec:example}
As shown in Fig.~2, we discuss four typical scenarios for wireless big data processing in this section, including mobile crowd sensing, proactive caching, cooperative vehicular networking, and fog computing, which all require incentive mechanism design.\par
\begin{figure}[!tbp]
\centering
\includegraphics[width=6.5in]{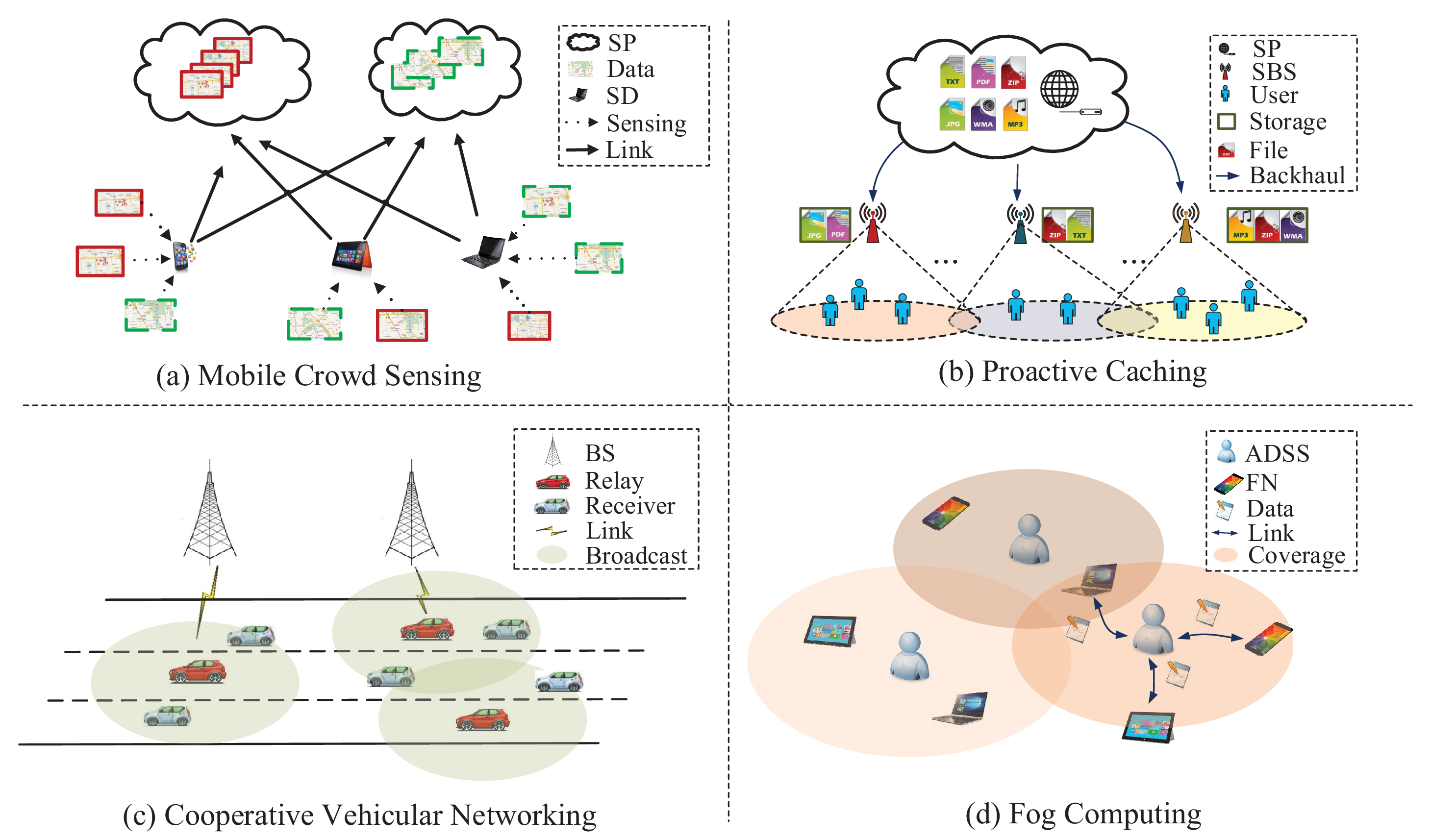}
\vspace{-5mm}
\caption{Typical scenarios in wireless big data processing with incentive mechanism design problems.}
\label{fig:fig2}
\vspace{-5mm}
\end{figure}
\subsection{Mobile Crowd Sensing}
Thanks to the development of mobile crowd sensing~\cite{Sensing} as in Fig.~2(a), Internet SPs can collect sensing data, such as locations, wireless channel conditions, environment temperatures, and human health conditions from sensors in mobile devices, personal laptops, and on-board computers. With statistics and machine learning approaches, the SPs can extract typical features from the collected data and provide precise user-centric services on location-based applications, healthcare, environment monitoring, and social networking. The sensing devices~(SDs) of private users can be used to sense data for the SPs only with proper payments. Additionally, each SD can only sense limited data due to limitations on the sensing capacity. Therefore, the SPs have to compete for the sensing resources on the SDs. Game theory can be used to model the competition among the SPs, the payment negotiations and the sensing resource allocation. Since the quality of statistical processing highly depends on the amount of data and their heterogeneity, each SP usually needs to find a large number of SDs. Thus, the game between the SPs and the SDs is a large-scale one.\par
\subsection{Proactive Caching}
Ultra-dense wireless infrastructure deployments, such as small-cell base stations~(SBSs), are considered to be a promising approach to deal with the rapidly increasing wireless traffic by achieving high-density spatial reuse of communication resources. However, the backhaul link capacity between the SBSs and the SPs in the core networks is the major obstruction to provide satisfactory download speed. With proactive caching~\cite{Caching} as shown in Fig.~2(b), the SPs can forecast mobile users' file demands and proactively cache popular content in the storage of the SBSs during off-peak times. Then, duplicate backhaul transmission can be reduced and heavy traffic can be alleviated during peak times. Correspondingly, each mobile user can experience a better quality of service~(QoS) for the desired content. In this case, the SPs need to rent storage on the SBSs and negotiate the corresponding payments. When the communication network is ultra-densed, there may exist hundreds of the SBSs in a local area. Thus, a large-scale game is needed to study this problem. Fast convergence of the game is usually required, which can provide the SPs with more opportunities to respond to the realtime popularity of the content through more frequent file replacements. \par
\subsection{Cooperative Vehicular Networking}
To support various applications for the vehicular users, the widely deployed Long Term Evolution (LTE)
networks are incorporating an integrated system to facilitate vehicle-to-vehicle~(V2V) and vehicle-to-infrastructure~(V2I) communications with low
latency and high reliability~(LLHR)~\cite{Vehicular}. In addition to cellular data transmission from the infrastructure~(e.g., BSs), each vehicle can achieve information from the nearby vehicles through device-to-device~(D2D) communications. As shown in Fig.~2(c), a number of vehicles can work as relays and forward data for other vehicles. For example, in downlink transmission, the BSs can first select some vehicles as relays and transmit a bundle of multimedia content to them. Then, these relays can broadcast the content to the vehicles nearby. Naturally, the BSs need to pay for the assistance of the vehicles. To meet LLHR requirements in consideration of the mobility of vehicles, the payment negotiation must converge quickly. \par
\subsection{Fog Computing}
The development of DCs can provide enhanced computing power and storage services for authorized data
service subscribers~(ADSSs). However, DCs are sometimes built far away from SPs, therefore, some real-time services cannot be hosted on the DCs. Then, a number of flexible
computing devices close
to the ADSSs, namely fog nodes~(FNs)~\cite{Fog}, such as smart phones and personal laptops, can help process the ADSSs' tasks. Due to their proximity to the ADSSs, the
FNs can provide data services with low latency and fast response.  However, the ADSSs must engage a great number of FNs to accomplish the same tasks since the computation power of each FN is much lower than that of a server in the DCs. Thus, the incentive mechanisms between the ADSSs and the FNs must be scalable to handle the fragmental computing resource allocation in fog computing.\par
\section{Fast Convergent Game}
\label{sec:hierarchicalgame}
To deal with the incentive design problems in widespread large-scale wireless hierarchies, we develop a generalized, rapidly convergent, and scalable hierarchical game. We first provide the basic form of the hierarchical game to deal with the networks with one controller and multiple agents. Then, extended forms are provided with embedding ADMM into the hierarchical game, and finally we discuss their properties to indicate why it is suitable for wireless big data processing.\par
\begin{figure}[!tbp]
\centering
\includegraphics[width=6in]{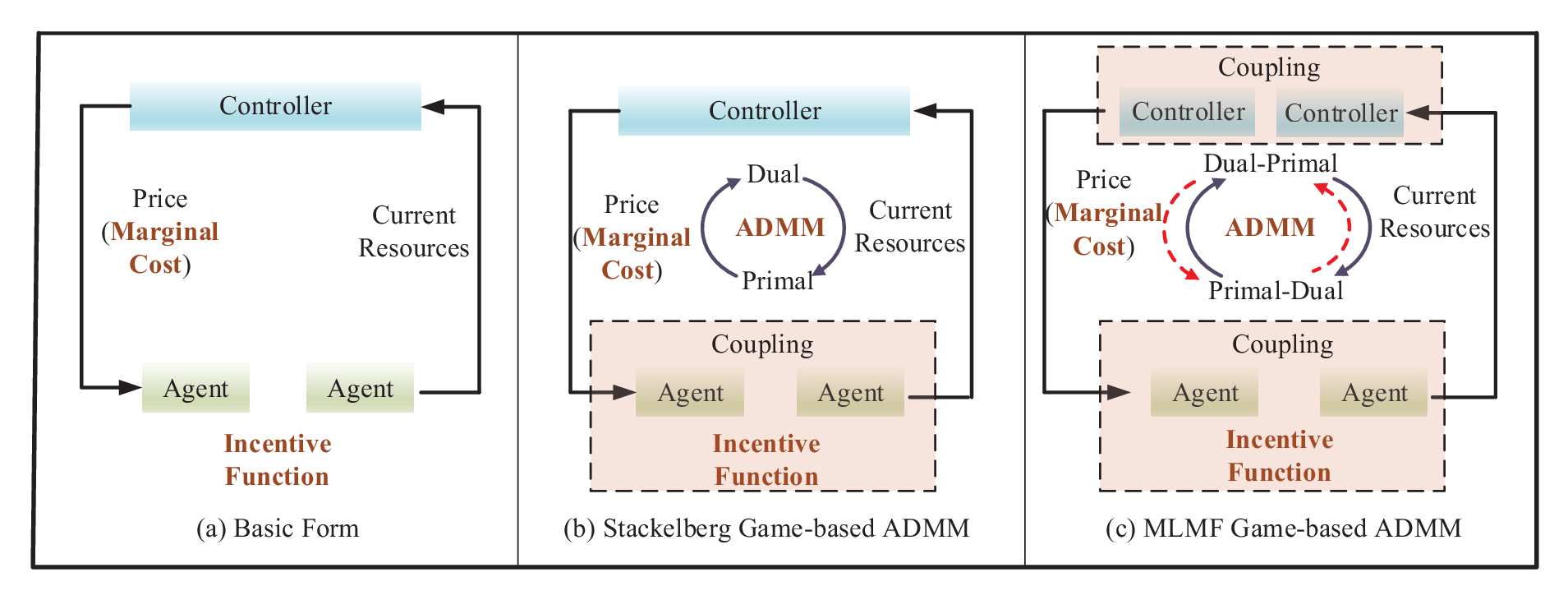}
\vspace{-5mm}
\caption{Three forms of the hierarchical game.}
\label{fig:fig3}
\vspace{-5mm}
\end{figure}
\subsection{Basic Form}
The basic form of the hierarchical game is designed to handle networks with one controller and multiple agents~\cite{Basic}, as shown in Fig.~3(a), where the agents are not congested with each other. The controller's task is quantified as the minimization of a decomposed function as $\sum_{i} f_i(x_i)$, where the function, $f_i$, depends only on the resources, $x_i$, rented from the $i^{th}$ agent. Each agent $i$'s utility is modeled as the payment from the controller, $\theta_i x_i$, minus the cost, $g_i(x_i)$, of providing resources for the controller's task. $\theta_i$ represents the payment per unit resource, i.e., the price. Then, the controller will adjust payments iteratively to the agents in order to incentivize the agents and provide enough resources for the controller's task. With a given price, each agent will provide the resources to maximize its utility, i.e., to achieve the highest payment with the lowest cost.
The iteration for the price and resource adjustments is operated in parallel at different agents. With two adaptations of the iterative process, the hierarchical game can guarantee to reach a Stackelberg equilibrium.
\begin{itemize}
\item \textbf{Incentive Function:} Besides considering the agent's own utility, each agent also considers the controller's task simultaneously. Specifically, the utility function of each agent~$i$ is replaced by an incentive function at each step, which is given as below
    $$\Phi_i(\theta_i,x_i)=-f_i(x_i)+\left(\theta_ix_i-g_i(x_i)\right),$$
    where the first term, $-f_i(x_i)$, is part of the controller's objective function to measure the accomplishment of the controller's task at agent $i$, and the second term is agent $i$'s utility. The physical meaning of the incentive function can be interpreted as follows.  The controller asks the agent $i$ to process the controller's task, $-f_i(x_i)$, with the payment, $\theta_ix_i$. Agent~$i$ can simultaneously process its own task, $-g_i(x_i)$. It can be found that the function is an intuitive description for the term ``incentive", i.e., the controller asks each agent to help the controller process the data with payment.
\item \textbf{Marginal Cost:} The price adjustment to agent $i$ by the controller at step $p+1$ equals agent $i$'s marginal cost at the current step, as follows:
    $$ \theta^{(p+1)}=\nabla_{x_i} g_i\left(x_i^{(p)}\right), $$
    where $\theta^{(p+1)}$ is the price to agent $i$ at step $p+1$, $x_i^{(p)}$ represents the resources provided by agent $i$ at step $p$, and $\nabla_{x_i}$ is the first order derivative. The term ``marginal" represents the temporary loss per unit resource of agent $j$'s utility, which can be mathematically calculated as the first order derivative.
\end{itemize}
\par
The discussion on the specific forms for the above functions will be provided in Section~\ref{sec:properties}.\par
\subsection{Extended Forms}
The basic form can be extended to two complicated forms through embedding ADMM into the game to cope with scenarios with more complicated restrictions.
\begin{itemize}
\item \textbf{Stackelberg Game-based ADMM~\cite{Extend1}:} When the network consists of one controller and multiple conflicted agents, i.e., with some coupling constraints on the resources from different agents, the above basic form cannot be applied since the adjustment of the resources on one agent will influence others. To deal with the conflicts among the agents, we embed the ADMM process~\cite{ADMM} as an inner loop into the basic form as shown in Fig.~3(b). At each step with given prices, the agents can cooperatively reach a consensus solution with an iterative primal-dual update as in the ADMM. A dual variable is updated by the controller to handle the coupling constraints. The convergence of the ADMM ensures the inner loop to reach a solution that the coupling constraints are satisfied. Then our designed price and resource adjustments in the outer loop are guaranteed to reach a Stackelberg equilibrium.
\item \textbf{Multi-Leader Multi-Follower~(MLMF) Game-based ADMM~\cite{Extend2}:} When the network includes multiple controllers and multiple agents, as in Fig.~3(c), congestion among controllers must also be considered. Similarly, an ADMM-like process is introduced as an inner loop for the basic form. Different from a single primal-dual update in the Stackelberg game-based ADMM, more dual variables are included for both the controllers and the agents to cope with coupling constraints on both sides. Specifically, each controller has a dual variable to help cope with the conflicts among the agents. On the other hand, a dual update is also included on each agent to handle the controllers' conflicts. This extended form can reach a hierarchical social optimum, where the sum of the objective functions for all controllers is minimized, as the game among agents reach a Nash equilibrium.
\end{itemize}
\subsection{Properties}
\label{sec:properties}
Properties concerning convergence conditions, convergence speed, and scalability are summarized here to corroborate the potential of our hierarchical game in wireless big data processing. All statements below have analytical proofs and derivations that can be found~\cite{Extend1}.
\begin{itemize}
\item \textbf{Convergence Conditions:} Our hierarchical game, in either the basic form or the two extended forms, is guaranteed to converge when the following conditions are simultaneously satisfied:
\begin{itemize}
\item The objective function for each controller can be decomposed as a sum of independent convex functions, $\sum f_i(x_i)$. The value of each convex function depends only on at most one agent's provided resources.
\item The cost function for each agent is convex in the basic form and can be decomposed as a sum of independent convex functions in the extended forms. The value of each convex function depends only on the agent's resources for at most one controller.
\item The conflicts among the agents are described by a group of linear constraints on the agents' resources. Each constraint includes the resources from different agents.
\item The conflicts among the controllers are also described by a group of linear constraints on the agents' resources. Each constraint includes the resources required by various controllers.
\end{itemize}
\item \textbf{Convergence Speed:} Our hierarchical game can converge at a linear speed. Specifically, to measure the accuracy, we use the Euclidean distance, $\varepsilon$, between the value of the controllers' objective functions achieved in our game and the optimal ones. Then, the iteration for payment negotiation and payment adjustments between each controller and each agent is upper bounded by $O(\log(1/\varepsilon))$, i.e., a linear convergence speed, when the following additional conditions are also satisfied:
\begin{itemize}
\item Any convex part in each controller's objective function is strongly convex, which indicates that the first order derivative of a function cannot change too slow.
\item Any convex part in each agent's cost function satisfies a uniform Lipschitz gradient condition, which usually implies that the first order derivative of a function cannot change too fast.
\end{itemize}
    The above conditions clarify the generality and feasibility of our game. We emphasize that all conditions are common and many general functions, such as log functions and quadratic functions, satisfy these conditions. Similar conditions can also be found in convex optimization approaches~\cite{CVX}.
\item \textbf{Scalability:} It can be directly shown in the basic form that the convergence speed will not change with the variation of the network size since the objective function of the controller is assumed to be decoupled without coupling constraints. As for the extended forms, the inner loop with ADMM can help satisfy the coupling constraints for both the controllers and the agents. It can be shown that the convergence time in the outer loop for the price negotiations and resource adjustments in the extended forms is independent of the network size.
\end{itemize}
\par
For performance illustration, we illustrate the performance of the MLMF game-based ADMM in mobile crowd sensing~\cite{Extend2}. The other forms and more applications can be found in~~\cite{Fog}, \cite{Basic}, and ~\cite{Extend1}, which can provide the similar illustrations. The objective functions of the controllers are selected as typical functions in sensing problems, such as the quadratic functions for model fitting problems. Shannon capacity, a log function~(without interference), is selected as the cost function for each agent. \par
\begin{figure}[!tbp]
\centering
\includegraphics[width=6.5in]{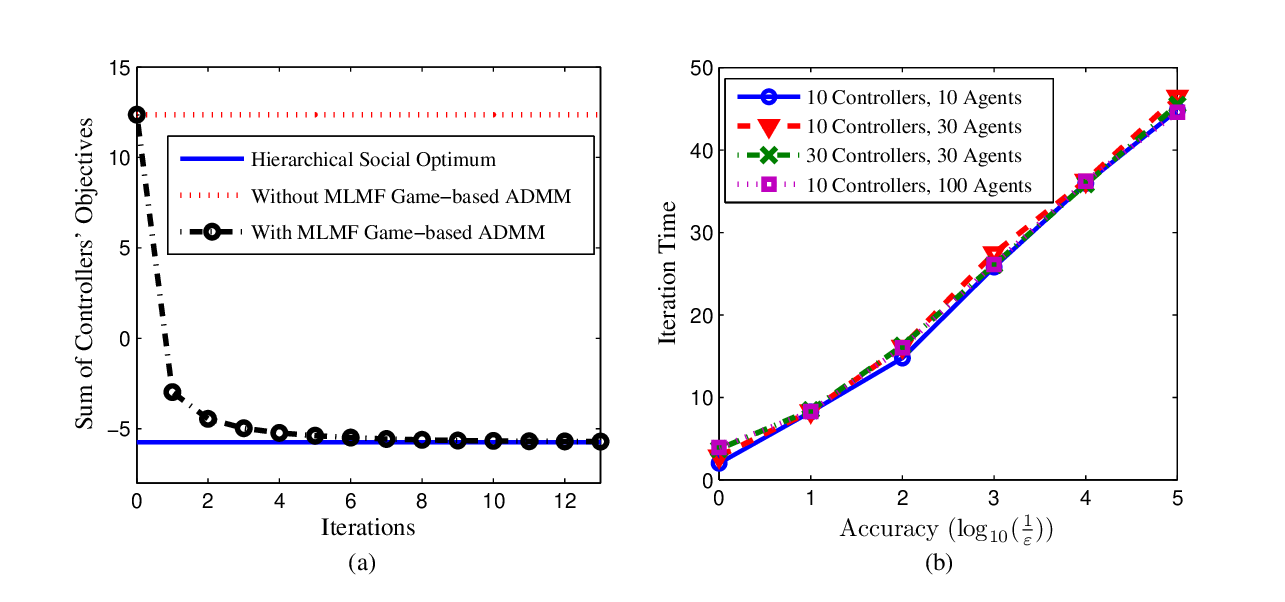}
\vspace{-10mm}
\caption{Performance illustration in mobile crowd sensing.}
\label{fig:fig4}
\vspace{-5mm}
\end{figure}
Fig. 4(a) compares the sum of the controllers' objective functions of
the MLMF game-based ADMM with that of the hierarchical
social optimum and without the MLMF game-based ADMM.
From the figure, the MLMF game-based ADMM converges
into a hierarchical social optimum within only 13 iterations when the network consists of 10 controllers and 100 agents. \par
Fig. 4(b) shows the required number of iterations versus
the accuracy, $\varepsilon$, for networks with different sizes. From the figure, the relation between the required number of iterations and $\log(1/\varepsilon)$ is linear, which verifies the linear convergence and the scalability of our game. For example, when~$\varepsilon=10^{-3}$, the required
number of iterations is always about 26, and is obviously independent
of the numbers of controllers or agents.\par

\section{Future Work}
In this section, we outline several possible future research directions in this emerging area.
\label{sec:future}
\subsection{More General Objective Functions}
As mentioned in Section~\ref{sec:properties}, the rapid convergence and scalability are true only when the objective functions for both the controllers and the agents are decomposable convex functions. For more generalized coupled objective functions and convex sets, the corresponding proofs and simulation tests have not yet been provided. Besides case-by-case analysis and simulation, possible compromise approaches are pre-decomposing the objective functions and linearizing the constraints with approximation before the game formulation. \par

\subsection{Widespread Scenarios}
The scenarios that have already implemented our approaches have been summarized as C-RAN in~\cite{Basic}, mobile crowd sensing in~\cite{Extend1} and~\cite{Extend2}, and fog computing in~\cite{Fog}. Besides the possible scenarios as mentioned in Section~\ref{sec:example}, our hierarchical game can help design incentive mechanisms for hierarchical structures, such as resource slicing in the WCNs, IoT applications, cloud computing, data center, smart grids, etc. For example, in 5G virtualized networks, a hierarchal game can be adopted by SPs to
induce the network operators to share their bundled network resources. \par

\subsection{Other Game Theoretic Methods}
Our hierarchical game has potential to be extended to other games with a hierarchy.
For example, in the moral-hazard contract game with one controller and multiple agents,
the utility for each risk-averse agent is a concave function of the wage.
The controller can design the wage as the marginal cost to preclude full insurance for each agent rapidly.
In addition, to guarantee each agent feeding back the truthful marginal cost, second price auction can be embedded for price adjustments. Furthermore, considering local interactions between various agents in large-scale networks, the idea of mean-field dynamics in game theory can also be combined with our hierarchical game.\par

\section{Conclusion}
\label{sec:conclusion}
In this article, game theoretic approaches have been introduced as incentive mechanisms to offload the big data processing into a large number of agents. Four typical scenarios, mobile crowd sensing, proactive caching, cooperative vehicular networking, and fog computing, have been discussed to illustrate the incentive mechanism design problems in wireless hierarchies. We have proposed a hierarchical game and provided three specific forms, all of which can be proved to converge when the objectives are convex functions. Rapid convergence and scalability are guaranteed, which supports their use in the large-size and real-time wireless big data processing. The specific example has been provided to demonstrate the convergence properties of our hierarchical game. Possible future research directions have also been outlined.\par

\section{Acknowledgement}
This work is partially supported by National Science and Technology Major Project of China under grant number 2016ZX03001017 and the National
Nature Science Foundation of China under grant number 61625101 and 61511130085, and is also partially supported by US NSF CNS-1717454, CNS-1731424, CNS-1702850, CNS-1646607, ECCS-1547201, CMMI-1434789, CNS-1443917, and ECCS-1405121.\par



\begin{IEEEbiographynophoto}
{Zijie Zheng} received the B.S. degree in electronic engineering from Peking University, China, in 2014, where he is currently pursuing the Ph.D. degree with the School of Electrical Engineering and Computer Science. His current research interests include game theory in 5G networks, wireless powered networks, mobile social networks, and wireless big data.
\end{IEEEbiographynophoto}

\begin{IEEEbiographynophoto}
{Lingyang Song} is a Professor in the School of Electronics Engineering and Computer Science, Peking University, China. His main research interests include MIMO, cognitive and cooperative communications, security, and big data. He was a recipient of the IEEE Leonard G. Abraham Prize in 2016 and IEEE Asia Pacific Young Researcher Award in 2012. He is currently on the Editorial Board of the IEEE Transactions on Wireless Communications.
\end{IEEEbiographynophoto}

\begin{IEEEbiographynophoto}
{Zhu Han} is a Professor in the Electrical and Computer Engineering Department as well as in the Computer Science Department at the University of Houston, Texas. His research interests include wireless communications and networking, game theory, big data analysis, and smart grid. Dr. Han received IEEE Leonard G. Abraham Prize in the field of Communications Systems (best paper award in IEEE JSAC) in 2016, and several best paper awards in IEEE conferences.
\end{IEEEbiographynophoto}

\begin{IEEEbiographynophoto}
{Geoffrey Ye Li} is a Professor with Georgia Tech. His general research is in signal processing and machine learning for wireless communications. He has published over 400 articles with around 30,000 citations and is listed as a Highly-Cited Researcher by Thomson Reuters. He has been an IEEE Fellow since 2006. He won 2010 Stephen O. Rice Prize Paper Award and 2017 Award for Advances in Communication from the IEEE ComSoc.
\end{IEEEbiographynophoto}

\begin{IEEEbiographynophoto}
{H. Vincent Poor} is the Michael Henry Strater University Professor of Electrical Engineering at Princeton University. His research interests include information theory and signal processing, with applications in wireless networks, energy systems, and related fields. An IEEE Fellow, he is also a member of the NAE and NAS and a foreign member of the Royal Society. Recent recognition of his work includes the 2017 IEEE Alexander Graham Bell Medal and honorary doctorates from several universities.
\end{IEEEbiographynophoto}



\begin{thebibliography}{15}
\bibitem{Cisco}
Cisco White Paper, ``Visual networking index: global mobile data traffic forecast update, 2016-2021", Mar. 2017.

\bibitem{Bigdatamag2}
X. Zhang, Z. Yi, Z. Yan, G. Min, W. Wang, A. Elmokashfi, S. Maharjan, and Y. Zhang, ``Social computing for mobile big data", \emph{IEEE Comput. Mag.}, vol.~49, no.~9, pp.~86-90, Sept. 2016.


\bibitem{Bigdatamag}
S. Bi, R. Zhang, Z. Ding and S. Cui, ``Wireless communications in the era of big data," \emph{IEEE Commun. Mag.}, vol. 53, no. 10, pp. 190-199, Oct. 2015.

\bibitem{Architecture}
K. Wang, Y. Wang, X. Hu, Y. Sun, D. Deng, A. Vinel, and Y. Zhang, ``Wireless big data computing in smart grid", \emph{IEEE Wireless Commun.}, vol.~24, no.~2, pp.~58-64, Apr.~2017.

\bibitem{Hadoop}
\emph{Apache Hadoop}. Available: \url{http://hadoop.apache.org/}

\bibitem{ADMM}
S. Boyd, N. Parikh, E. Chu, B. Peleato, and J. Eckstein, ``Distributed optimization
and statistical learning via the alternating direction method of multipliers," \emph{Found. and Trends in Mach. Learning}, vol. 3, no. 1, pp.
1-122, Nov. 2010.

\bibitem{Game}
Z. Han, D. Niyato, W. Saad, T. Basar, and A. Hjorungnes, \emph{Game
Theory in Wireless and Communication Networks: Theory, Models and
Applications}, Cambridge, U.K.,: Cambridge Univ., 2011.


\bibitem{Sensing}
B. Di, T. Wang, L. Song, and Z. Han, ``Collaborative smartphone sensing using overlapping coalition formation games," \emph{IEEE Trans. Mobile Computing}, vol. 16, no. 1, pp. 30-43, Jan. 2017.


\bibitem{Caching}
Z. Hu, Z. Zheng, T. Wang, L. Song, and X. Li, ``Caching as a service: small-cell caching mechanism design for service providers," \emph{IEEE Trans. Wireless Commun.}, vol. 15, no. 10, pp. 6992-7004, Oct. 2016.


\bibitem{Vehicular}
G. Araniti, C. Campolo, M. Condoluci, A. Iera, and A. Molinaro, ``LTE for vehicular networking: a survey," \emph{IEEE Commun. Mag.}, pp. 148-157, vol. 51, no. 5, May 2015.


\bibitem{Fog}
N. Raveendran, H. Zhang, Z. Zheng, L. Song, and Z. Han, ``Large-scale fog computing optimization using
equilibrium problem with equilibrium constraints", accepted by \emph{IEEE Global Commun. Conf. ~(GLOBECOM17)}, Singapore, Dec. 2017.

\bibitem{Basic}
Z. Zheng, L. Song, and Z. Han, ``Bridging the gap between big data and game theory:
a general hierarchical pricing framework", accepted by \emph{IEEE Int. Conf. Commun.~(ICC17)}, Paris, France, May 2017.

\bibitem{Extend1}
Z. Zheng, L. Song, and Z. Han, ``Bridge the gap between ADMM and
stackelberg game: incentive mechanism design for big data networks," \emph{
IEEE Signal Process. Lett.}, vol. 24, no. 2, pp. 191-195, Feb. 2017.

\bibitem{Extend2}
Z. Zheng, L. Song, Z. Han, G. Li, and V. Poor, ``Multi-leader multi-follower game-based ADMM for big data processing," \emph{IEEE Int. Workshop Signal Process. Advances Wireless Commun.~(SPAWC17)}, Sapporo, Japan, Jul. 2017.

\bibitem{CVX}
S. Boyd and L. Vandenberghe, \emph{Convex Optimization}, Cambridge, U.K.,:
Cambridge Univ., 2004.




\end{thebibliography}
\end{document}